\documentclass{emulateapj}
\usepackage{color}

\def\be{\begin{equation}}
\def\ee{\end{equation}}
\def\mbh{M_{\bullet}}

\def\mst{M_{*,\rm sph}}
\def\msun{M_{\odot}}

\def\kms{{\rm \,km\,s^{-1}}}
\def\rhobh{\rho_{\bullet}}

\def\dex{\rm\, dex}
\def\keV{\rm\, keV}

\begin{document}

\title{The cosmic evolution of massive black holes and galaxy spheroids: 
global constraints at redshift $z\la 1.2$
}
\author{Xiaoxia Zhang$^1$, Youjun Lu$^1$ \& Qingjuan Yu$^{2}$}
\affil{$^1$~National Astronomical Observatories, Chinese Academy of Sciences,
Beijing, 100012, China; luyj@nao.cas.cn \\
$^2$~Kavli Institute for Astronomy and Astrophysics,
Peking University, Beijing, 100871, China; yuqj@pku.edu.cn
}

\begin{abstract}

We study the observational constraints on the cosmic evolution of the
relationships between the massive black hole (MBH) mass ($\mbh$) and
the stellar mass ($\mst$; or velocity dispersion $\sigma$) of the host
galaxy/spheroid. Assuming that the $\mbh-\mst$ (or $\mbh-\sigma$)
relation evolves with redshift as $\propto (1+z)^{\Gamma}$, the MBH
mass density can be obtained from either the observationally
determined galaxy stellar mass functions or velocity dispersion
distribution functions over redshift $z\sim 0-1.2$ for any given
$\Gamma$. The MBH mass density at different redshifts can also be
inferred from the luminosity function of QSOs/AGNs provided known
radiative efficiency $\epsilon$. By matching the MBH density inferred
from galaxies to that obtained from QSOs/AGNs, we find that
$\Gamma=0.64^{+0.27}_{-0.29}$ for the $\mbh-\mst$ relation and
$\Gamma=-0.21^{+0.28}_{-0.33}$ for the $\mbh-\sigma$ relation, and
$\epsilon=0.11^{+0.04}_{-0.03}$.  Our results suggest that the MBH
mass growth precedes the bulge mass growth but the galaxy velocity
dispersion does not increase with the mass growth of the bulge after
the quench of nuclear activity, which is roughly consistent with the
two-phase galaxy formation scenario proposed by \citet{Oser12} in
which a galaxy roughly double its masses after $z=1$ due to accretion
and minor mergers while its velocity dispersion drops slightly.

\end{abstract}


\section{INTRODUCTION}

The masses of the massive black holes (MBH; $\mbh$) are tightly
correlated with the properties of the spheroidal components of their
host galaxies, such as the velocity dispersion $\sigma$
\citep[e.g.,][]{Gultekin09, Gebhardt00, FM00, Tremaine02, Graham11},
the luminosity $L_{\rm sph}$ \citep[e.g.,][]{Kormendy95,McLure01}, and
the stellar mass $M_{\rm *,sph}$ \citep{Magorrian98, McLure02,
Marconi03, Haring04}, which suggests a strong link between the growth
of MBHs and the evolution of their host galaxies (or particularly the
galaxy spheroids).  Feedback from the nuclear activities is proposed
to be responsible for the establishment of these relationships, either
through momentum or energy driven winds to self-regulate the MBH
growth \citep[e.g.,][]{SR98, Fabian99, King03, WL03, Murray05,
diMatteo05, Croton06, Bower06, Somerville08}. However, the detailed
physics on how the feedback mechanism processes is still not clear.

Observational determination of the cosmic evolution of the relations
between $\mbh$ and $\mst$ (or $\sigma$ or $L_{\rm sph}$) may reveal
important clues to the origin of these relations and put constraints
on feedback mechanisms \citep[][]{Shields03, Peng06, Woo06, Treu07,
Salviander07, Woo08, Alexander08, Somerville09, KK10, Merloni10,
Sarria10, Bennert10, Bennert11, Cisternas11, SW11, Portinari12,
LCL12}.  A number of studies have shown that the $\mbh-\mst$ relation
in active galactic nuclei (AGNs) evolves with redshift (or cosmic
time), $\propto (1+z)^{\Gamma}$, where $\Gamma\sim 0.68-2.1$
\citep[e.g.,][]{Merloni10,Bennert11}. There are also tentative
observational evidences suggesting that the $\mbh-\sigma$ relation may
also evolve with redshift \citep[e.g.,][]{Woo06,Woo08}. In most of
those studies, the MBH masses are derived by adopting the virial mass
estimators, which are based on the mass estimates of several dozen
MBHs in nearby AGNs through the reverberation mapping technique and a
calibration of those masses to the $\mbh-\sigma$ relation obtained for
nearby normal galaxies \citep[e.g.,][]{Onken04, Graham11}.  However,
the MBH mass estimated from the virial mass estimators may suffer from
some systematic biases due to various reasons
\citep[e.g.,][]{Krolik01,Collin06, NM10,KZ11,Graham11}; moreover, the
MBHs in AGNs are still growing rapidly, unlike those in nearby
quiescent galaxies. Therefore, it is not yet clear whether the cosmic
evolution of the $\mbh-\mst$ relation and the $\mbh-\sigma$ relation
found for AGNs are biased or not and whether the relations for normal
galaxies has a similar cosmic evolution as those for AGNs.

  In this paper, we adopt an alternative way to investigate the cosmic
evolution of the $\mbh-\mst$ relation and the $\mbh-\sigma$ relation
in normal galaxies by matching the MBH mass density inferred from
normal galaxies at different redshifts with that inferred from AGNs.
The evolution of the $\mbh-\mst$ and $\mbh-\sigma$ relations for
normal galaxies is assumed to follow a simple power-law form, i.e.,
$\propto (1+z)^{\Gamma}$. In Section~\ref{sec:den_gal}, we estimate
the evolution of the MBH mass density in normal galaxies by the two
ways: (i) using the stellar mass functions (SMFs) of galaxies
determined by recent observations over the redshift range $z\sim
0-1.2$ \citep[e.g.,][]{Bernardi10, Ilbert10}, and (ii) using the
velocity dispersion functions (VDFs) of galaxies at $z\sim 0-1.5$
\citep{Bernardi10,Bezanson11}. The evolution of the estimated MBH
densities depends on the parameter $\Gamma$.  According to the simple
\citet{Soltan82} argument, the MBH mass density evolution can also be
derived from the AGN luminosity functions (LFs), as shown in
Section~\ref{sec:den_agn}, where the parameter $\Gamma$ is not
involved. By matching the MBH density evolution inferred from
properties of normal galaxies to that inferred from AGNs, the cosmic
evolution of the $\mbh-\mst$ (or $\mbh-\sigma$) relation is then
constrained in Section~\ref{sec:fit}.  Discussion and conclusions are
given in Sections~\ref{sec:discussion} and \ref{sec:conclusion}.

Throughout the paper, we adopt the cosmological parameters
$H_0=70.5\kms$, $\Omega_{\Lambda}=0.726$, and $\Omega_{\rm M}=0.274$
\citep{Komatsu09}. The mass is in units of $M_\odot$ and the velocity
dispersion is in units of $\kms$. Given a physical variable $X$ (e.g.,
mass, logarithm of mass, velocity distribution), the $X$ function is
denoted by $n_X(X,z)$ so that $n_X(X,z)dX$ represents the comoving
number density of the objects (e.g., MBHs or galaxies) with variable
$X$ in the range $X\rightarrow X+dX$ at redshift $z$.
  
\section{The mass density evolution of massive black holes }
\label{sec:den_gal}

\subsection{The mass density of massive black holes inferred from
stellar mass functions} \label{subsec:smf}

The mass of a MBH in the center of a nearby ($z\sim 0$) normal galaxy
can be estimated through the $\mbh-\mst$ relation at $z=0$, i.e.,
\begin{eqnarray}
\left<\log\mbh\right>(\mst;z=0) & = & (8.20\pm 0.10)+ (1.12\pm0.06) \nonumber \\
                                &   & \times (\log\mst-11), 
\label{eq:mbhms}
\end{eqnarray}
where $\left<\log\mbh\right>(\mst;z=0)$ is the mean value of the
logarithmic MBH masses for galaxies with spheroidal mass $\mst$, and
the intrinsic scatter of $\log\mbh(\mst;z=0)$ around this mean value
is $0.3\dex$ \citep[][see also
\citealt{McLure02}]{Haring04}.
Currently, there is still no consensus on the $\mbh-\mst$ relation for
galaxies at high redshift. In general, the $\mbh-\mst$ relation may
evolve with redshift and the evolution may be simplified by 
\begin{eqnarray}
\left<\log\mbh\right>(\mst;z) & = & \left<\log\mbh\right>(\mst;z=0) \nonumber\\
                              &   & +\Gamma\log(1+z) 
\label{eq:mbhmstar}
\end{eqnarray}
as assumed in a number of previous works \citep[e.g.,][]{Merloni10,
Bennert11}, where the parameter $\Gamma$ describes the significance of
the evolution, and the intrinsic scatter of the relations is assumed
not to evolve with redshift.

The mass function of MBHs (BHMF) at redshift $z$ may be estimated by
adopting the $\mbh-\mst$ relation (eq.~\ref{eq:mbhmstar}) and the SMF
of spheroids at that redshift if the SMF can be observationally
determined, i.e., 
\begin{eqnarray}
n_{\log\mbh}(\log\mbh,z) & = &\int n_{\log\mst}(\log\mst,z)\times \nonumber \\
                         &   &P(\log\mbh;\left<\log\mbh \right>)  d\log\mst,
\label{eq:bhmf}
\end{eqnarray}
where $n_{\log\mst}(\log\mst,z)$ is the SMF of spheroids at redshift
$z$, $P$ is the probability density function of $\log\mbh$ around
$\left< \log\mbh \right>$ and is assumed to be normally distributed
with a dispersion of $0.3\dex$.  The SMF of spheroids at redshift $z$
can be estimated by using the SMFs for galaxies with different
morphological types and the bulge-to-total mass ratios $B/T$, i.e.,
\begin{eqnarray}
n_{\log\mst}(\log\mst,z) & = & \sum_i n^i_{\log\mst}(\log\mst,z) \nonumber 
\end{eqnarray}
\be
                          =  \sum_i f^i(M_{*,\rm tot},z)  
                             n_{\log M_{*,\rm tot}}(\log M_{*,\rm tot},z)
\frac{d\log M_{*,\rm tot}}{d\log\mst},
\label{eq:smfs}
\ee
where $M_{*,\rm tot}=\mst/({\it B/T})$, $n_{\log M_{*,\rm tot}} (\log
M_{*,\rm tot},z)$ is the SMF for all galaxies,
$n^i_{\log\mst}(\log\mst,z)$ is the SMF for the spheroids of those
galaxies with Hubble type $i$, $f^i(M_{*,\rm tot},z)$ is the fraction
of galaxies with Hubble type $i$ to all galaxies, and the summation is
over galaxy morphological types from E, S0, Sa-Sb, Sc-Sd to Irr.

The SMFs for galaxies at $z\sim 0$ with different morphological types
have been obtained from the Sloan Digital Sky Survey (SDSS;
\citealt[][see Table B2 therein]{Bernardi10}). And the bulge-to-total
mass ratio $B/T$ has been given by \citet[][see Table 1]{Weinzirl09}
for more than a hundred nearby galaxies with different Hubble types.
In summary, the bulge-to-total mass ratios are $1$, $0.28\pm 0.02$,
$0.46\pm 0.05$, $0.35\pm 0.10$, $0.22\pm 0.08$, $0.15\pm 0.05$, and
$0$ for E, S0, Sa, Sab, Sb, Sc-Sd, and Irr, respectively.\footnote{
\citet{GW08} obtain the ratio of K-band bulge
luminosity/flux to disk luminosity/flux for a large number of disk galaxies in
which the dust extinction is considered and corrected. Assuming
a constant mass-to-light ratio, we also alternatively adopt the $B/D$
ratios inferred from \citet{GW08} in our analysis, i.e., $\log(B/D)$
= -0.54, -0.34, -0.54, -0.6, and -1.2 for S0, Sa, Sab, Sb, and Scd,
respectively, and obtain the constraints $(\Gamma,\epsilon)=(0.68\pm
0.24, 0.10\pm 0.01)$, which is consistent with that obtained by
adopting the $B/D$ ratios given by \citet{Weinzirl09}.} Adopting these
observations, the SMF of spheroids and the BHMF at $z=0$ can be
estimated. 

For galaxies at redshift $z\sim 0.2-1.2$, the total SMFs have been
obtained by \citet{Ilbert10}, \citet{PG08}, \citet{Fontana06}, and
\citet{Borch06}. These SMFs are usually obtained from deep surveys
with small sky coverage and may suffer the cosmic variance. To avoid
the cosmic variance, we adopt the average total SMFs according to the
SMFs estimated in the above papers. In \citet{Ilbert10}, the SMFs
of quiescent early type galaxies (E+S0) at different redshifts are directly
obtained (see Table 2 therein); and we obtain the SMFs of all the galaxies by
summing the SMFs of early type galaxies and those of ``intermediate activity''
and ``high activity'' galaxies (see the definition of ``intermediate activity''
and ``high activity'' in \citealt{Ilbert10}, and see their Table~3). We assume 
that the fraction of the early-type galaxies at any given mass is the 
same as that given by \citet{Ilbert10} and hereafter adopt this fraction,
together with the average total SMFs, to calculate the MBH mass
function for early-type galaxies. The LFs of galaxies were estimated
by \citet{Zucca06} for four different spectral types, which roughly
correspond to the morphological types of E/S0, Sa-Sb, Sc-Sd, and Irr,
respectively. For each type of galaxies, the mass-to-light ratio can
be estimated through their average colors, for instance, $
\log(M_{*,\rm tot}/L_{\rm B})=-0.942+1.737(B-V)+0.15 \label{eq:mlr} $
for early-type galaxies by adopting the Salpeter initial mass function
and $ \log(M_{*,\rm tot}/L_{\rm B})=-0.942+1.737(B-V)- 0.10 $ for
late-type galaxies by adopting the Chaberier initial mass function
\citep{Bell01,Bell03,Bernardi10,Chabrier03}.  The $B-V$ colors for
different morphological types of galaxies are given by
\citet{Fukugita95}. The luminosity evolution can be corrected for each
type of galaxies according to \citet{Bell03}. The LF for galaxies with
different morphological types \citep{Zucca06} can thus be converted to
the SMFs. According to these SMFs, the relative abundance of different
late-type galaxies (Sa-Sb, Sc-Sd, and Irr) over $z\sim 0.2-1.2$ can be
obtained at any given $M_{*,\rm tot}$. The SMF of spheroids and the
BHMF can then be estimated if the $B/T$ is averagely the same for
galaxies with the same spectral type (and maybe correspondingly the
same morphological type) but at different redshifts. Consequently, the
mass density accreted onto MBHs with mass $>\mbh$ can be obtained by 
\be
\rhobh^{\rm
gal}(z;>\mbh)=\int^{\infty}_{\mbh}(\mbh'-\mbh) n_{\log\mbh'}(\log\mbh',z)
d\log\mbh',
\label{eq:den} 
\ee 
if mergers of MBHs do not significantly contribute to the MBH growth 
and the seeds of those MBHs are smaller than $\mbh$ [see Eq.~(29) in 
\citet{YL04} and Eq.~(35) in \citet{YT02}]. And this mass density can 
be directly matched by that inferred from QSOs/AGNs (see Section~\ref{sec:den_agn}).
Hereafter, we set the lower limit of the MBH mass in the above
integration to be $10^6\msun$ unless otherwise stated, as the smallest
mass for those nearby MBHs, of which the mass is well measured and
adopted to determine the $\mbh-\mst$ (or $\mbh-\sigma$) relation, is
$\sim 10^6\msun$. Given observationally well determined SMFs (or VDFs
below), the MBH mass density estimated from the normal galaxies
depends on the evolution parameter $\Gamma$ in the $\mbh-\mst$ (or
$\mbh-\sigma$) relation.

\subsection{The mass density of massive black holes inferred from velocity
dispersion distribution functions}\label{sec:msigm}

The mass of a MBH can also be estimated through the $\mbh-\sigma$ relation
at redshift $z=0$,
i.e.,
\begin{eqnarray}
 \left<\log\mbh\right>(\sigma;z=0)& = & (8.12\pm 0.08)+(4.24\pm0.41)  \nonumber \\
                                  &   & \times (\log\sigma-2.30), 
\label{eq:msigm}
\end{eqnarray}
with an intrinsic scatter of $0.44\dex$ \citep{Gultekin09}. Assuming an
evolutionary form similar to equation (\ref{eq:mbhmstar}), i.e.,
\begin{eqnarray}
 \left<\log\mbh\right>(\sigma;z)& = & \left<\log\mbh\right>(\sigma;z=0)  \nonumber \\
                                  &   & +\Gamma\log(1+z), 
\end{eqnarray}
the BHMF at redshift $z$ can be estimated through 
\be
n_{\log\mbh}(\log\mbh,z)=\int n^{\rm gal}_{\sigma}
(\sigma,z)  P(\log\mbh; \left<\log\mbh \right>) d\sigma,
\label{eq:phi}
\ee
where $n^{\rm gal}_{\sigma}(\sigma,z)$ is the galaxy VDF at redshift
$z$, $P$ is the probability density function of $\log\mbh$ around
$\left<\log\mbh \right>(\sigma;z)$ and is assumed to be normally
distributed with a dispersion of $0.44\dex$. The VDF for local
galaxies has been estimated from SDSS (\citealt[][see Table B4
therein]{Bernardi10}). At higher redshift $z\sim0.3-1.5$, the VDFs
have been obtained from UKIDSS (United Kingdom Infrared Telescope
Infrared Deep Sky Survey) Ultra-Deep Survey (UDS) and NEWFIRM (NOAO
Extremely Wide-Field Infrared Imager) Medium Band Survey (NMBS)
\citep{Bezanson11}. Adopting those VDFs, the BHMF and consequently the
mass density accreted onto MBHs with mass larger than $\mbh$ can also 
be estimated (see Eq.~\ref{eq:den}). 

\subsection{The mass density of massive black holes inferred from AGNs}
\label{sec:den_agn}

The MBH mass density at redshift $z$ can also be inferred from the LF
of AGNs according to the simple \citet{Soltan82} argument as MBHs
obtained their mass mainly through gas accretion
\citep[e.g.,][]{Salucci99, YT02, Marconi04, YL04, YL08, Shankar04,
Shankar09}, i.e.,
\begin{eqnarray}
\rho^{\rm AGN}_{\bullet}(z;>\mbh) & \simeq & \int^{\infty}_{z} dz \int dL_Y\int d C_Y
\frac{1-\epsilon}{\epsilon c^2} \times  \nonumber \\
 & & C_Y P(C_Y|L_Y)L_Y\phi(L_Y,z)
\left|\frac{dt}{dz}\right|,
\label{eq:soltan}
\end{eqnarray}
where $L_Y$ is the AGN $Y$-band luminosity, $\phi(L_Y,z)$  is the AGN
$Y$-band luminosity function, $\epsilon$ is the mass-to-energy
conversion efficiency, and $C_Y\equiv L_{\rm bol}/L_Y$ is the
bolometric correction (BC) for the $Y$ band, $L_{\rm bol}$ is the AGN
bolometric luminosity, and $P(C_Y|L_Y)dC_Y$ gives the probability of a
BC to be in a range $C_Y\rightarrow C_Y+dC_Y$ given a $L_Y$. The hard
X-ray LF of AGNs is adopted here because a significant number of
obscured AGNs can be detected only in the hard X-ray band while missed
in the optical surveys. Using the hard X-ray AGN LF and the
corresponding BC, the MBH mass density can be estimated given
a constant $\epsilon$. 

\begin{figure}
\epsscale{1.0}
\plotone{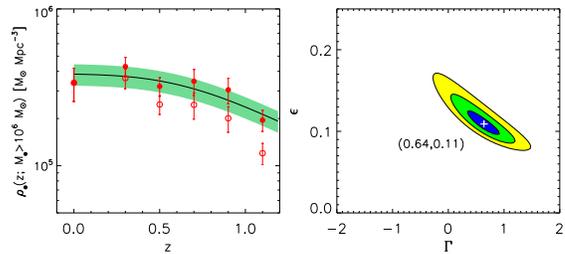}
\caption{
The cosmic evolution of the comoving MBH mass density. In the left
panel, the red solid circles with errorbars represent the mass densities
for those MBHs with mass $>10^6\msun$, estimated from the SMFs of 
normal galaxies by assuming an evolution with $\Gamma=0.64$ in the
$\mbh-\mst$ relation (see eq.~\ref{eq:mbhmstar}), which are best
matched by the solid line, while the red open circles represent the mass
densities estimated by assuming a universal $\mbh-\mst$ relation 
without evolution (i.e., $\Gamma=0$).  The solid line and the green shaded region
represent the MBH mass density inferred from the hard X-ray AGN LF by
assuming $\epsilon=0.11$ and its $1\sigma$ uncertainty.  The right
panel shows the best-fit parameter $(\Gamma,\epsilon)$ (denoted by
crosses) and the contours of the confidence levels for them, obtained
by matching the MBH densities inferred from the normal galaxies to
that inferred from AGNs.  The contours enclose the 68.3\%, 95.4\%, and
99.7\% confidence regions on the joint distribution of the two
parameters.
}
\label{fig:f1}
\end{figure}

The hard X-ray AGN LF has been estimated by a number of authors based
on the surveys by ASCA, Chandra, and XMM in the past decade
\citep[e.g.,][]{Ueda03, LaFranca05, Silverman08, Ebrero09, Yencho09,
Aird10}. In this paper, we adopt the latest 2-10$\keV$ X-ray AGN LF
obtained by \citet[][the LADE model in their Table~4]{Aird10} over the
redshift range $0\leq z\leq 3.5$ and extrapolate it to higher
redshift.\footnote{Since the Compton-thick AGNs are not included in
the hard X-ray AGN LF, which may be a fraction of $\sim 20\%$ of the
total AGN population \citep{Malizia09}, the MBH mass densities
inferred from the AGN LFs above may be underestimated by $20\%$ and thus
$\epsilon$ for the best match is underestimated by a factor of $\sim
1.2$.  } Adopting other versions of the 2-10$\keV$ AGN LF has little
effects on the results presented in the following section.  The BCs at
the hard X-ray band (2-10$\keV$, denoted as $C_X$) have been found to
be luminosity dependent and the BC at any given bolometric luminosity
has already been derived by \citet{Marconi04} and \citet{Hopkins07}.
To obtain the MBH mass density from the AGN LF, we need to estimate
the probability distribution function of $P(C_Y|L_Y)$.
\citet{Hopkins07} obtained the probability distribution function of
$C_Y$ as a function of $L_{\rm bol}$.  By adopting the same procedures
as those done in \citet{Hopkins07}, we fit the BCs by a log-normal
distribution with the following parameters\footnote{
We adopt the QSO spectral energy distribution (SED) model constructed 
by \citet{Hopkins07}. For details, the template SED consists of a power 
law in optical-UV band, i.e., $L_{\nu}\propto \nu^{\alpha}$ where $L_{\nu}$
is the energy radiated per unit time per unit frequency at frequency $\nu$, 
with $\alpha=-0.44$ 
for $1\mu{\rm m}< \lambda < 1300$\AA and $\alpha=-1.76$ from $1200$\AA~to $500$\AA.
At the wavelength longer than $\lambda>1\mu{\rm m}$, an infrared ``bump'' from
reprocessing of optical-UV-X-ray emission is adopted and truncated as a 
Rayleigh-Jeans tail of the
black-body emission ($\alpha=2$). The SED at energy high than $0.5$~keV is
determined also by a power law, with a slope of $-0.8$, and an
exponential cut-off at $500$~keV. For any given monochromatic luminosity
at $2500$\AA, the SED is renormalized to give the optical to X-ray slope
of $\alpha_{\rm ox} \equiv 0.384\log(L_{\tiny \nu({\rm 2500\AA})}/L_{\tiny \nu({\rm 2keV})})$ and
the points between $500$~eV and $500$\AA~are connected with a power law.
The value of $\alpha_{\rm ox}$ depends on luminosity as $\alpha_{\rm ox}=
-0.107\log (L_{\tiny \nu({\rm 2500\AA})}/{\rm erg~s^{-1}~Hz^{-1}})+1.739$. 
After the SED is constructed, we
integrate the SED to obtain the $2-10$~keV luminosity $L_{\rm X}$ and
the bolometric luminosity, and hence the BCs. We fit the BCs by a
double power-law function as shown in Equation (\ref{eq:bcx}).
}, 
\be
\langle \log C_X\rangle =\log\left[ c_1\left(\frac{L_{\rm X}}{10^{10}
L_\sun}\right)^{k_1}+c_2\left(\frac{L_{\rm X}}{10^{10} L_\sun}\right)^{k_2}\right],
\label{eq:bcx}
\ee
with $(c_1, k_1, c_2, k_2) = (26.75, 0.38, 7.44, 0.01)$, and the
scatter $\sigma_{\log C_X}=\sigma_1
\left(\frac{L_X}{10^9L_\sun}\right)^\beta+\sigma_2$ with
$(\sigma_1,\beta,\sigma_2)= (0.24, 0.06, -0.01)$. Similar to
\citet{YL08}, we then add a scatter of $0.15$~dex to include the X-ray
variabilities in AGNs according to \citet{Vasudevan07}. We adopt this
fitting formula of $C_X$ in the calculations of the MBH mass density.
In the calculations of the MBH mass density from AGNs for MBHs with
mass larger than $\mbh$, a lower limit is needed to set to the
luminosity. As the typical Eddington ratio for low-luminosity AGNs is
around $0.1$ \citep{Shen08} or $0.2$ \citep{Graham11}.
therefore, it is reasonable to set the lower limit $L_{\rm
bol}(\mbh)\simeq0.1L_{\rm EDD}(\mbh)$ or $0.2L_{\rm EDD}(\mbh)$. Our
calculations show that the difference in the lower limit does not lead
to significant difference in the results.

\begin{figure}
\epsscale{1.0}
\plotone{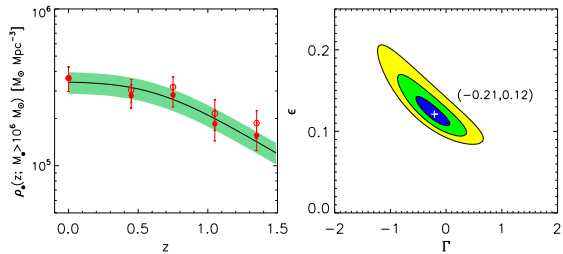}
\caption{ 
Legends are the same as those in Figure~\ref{fig:f1}, except that the
red solid (or red open) circles with errorbars are estimated from the VDFs 
of normal galaxies and by assuming an evolution with $\Gamma=-0.21$ 
(or no evolution with $\Gamma=0$) in the $\mbh-\sigma$ relation, and the red solid circles
are best matched by the solid line estimated from the AGN LF by assuming 
$\epsilon=0.12$. 
}                                                             
\label{fig:f2}                                                
\end {figure}                                                 

\section{Constraints on the cosmic evolution of the $\mbh-\mst$
relation and the $\mbh-\sigma$ relation } \label{sec:fit}

In this Section, we obtain constraints on the evolution of the
$\mbh-\mst$ (or $\mbh-\sigma$) relation by matching the MBH densities
inferred from normal galaxies with that from AGNs using the standard
$\chi^2$ statistics, i.e.,
\be
\chi^2\equiv \sum_{i}\frac{\left[\rho^{\rm gal}_{\bullet}(z_i,>\mbh)-\rho^{\rm
AGN}_{\bullet }(z_i,>\mbh)\right]^2}{{[\delta \rho^{\rm
gal}_{\bullet}(z_i,\mbh)}]^2+[\delta\rho^{\rm AGN}_{\bullet}(z_i,>\mbh)]^2},
\ee 
where $z_i$ is the $i$-th redshift bin and the summation is over all
the redshift bins, $\delta\rho^{\rm gal}_{\bullet}(z_i,>\mbh)$ and
$\delta\rho^{\rm AGN}_{\bullet}(z_i,>\mbh)$ are the uncertainties in
the MBH mass densities estimated from the normal galaxies and AGNs,
respectively, considering the $1\sigma$ errors in most of the fitting
parameters for the SMFs/VDFs and the AGN LFs, and the $\mbh-\mst$ and
$\mbh-\sigma$ relations. The errors in the normalizations of the
$\mbh-\mst$ and $\mbh-\sigma$ relations and the AGN LF only introduce
systematic shifts of all the estimates but do not lead to a change in
the shape of the cosmic evolution of the MBH density, therefore, we do
not include them in $\delta\rho^{\rm gal}_{\bullet}(z_i,>\mbh)$ and
$\delta\rho^{\rm AGN}_{\bullet}(z_i,>\mbh)$ in the $\chi^2$ fitting
here.

Figure~\ref{fig:f1} shows the mass densities of those MBHs with mass
$>10^6\msun$ estimated from the SMFs of normal galaxies, which are
best matched by the MBH densities estimated from the hard X-ray AGN
LFs. The corresponding parameters for the best match are
$(\Gamma,\epsilon)=(0.64^{+0.27}_{-0.29}, 0.11^{+0.02}_{-0.01})$ and
the errors are obtained for each parameter by marginalizing over the
other parameter. According to our calculations, a $\Gamma$ larger than
$1.6$ is excluded at the $3\sigma$ level. The value of $\Gamma=0.65$
suggests that the $\mbh-\mst$ relation evolves with redshift
positively, i.e., the relative positive offset of the $\mbh-\mst$
relation at redshift $z$ to that at local universe increases with
increasing redshift. Figure~\ref{fig:f2} shows the MBH mass densities
estimated from the VDFs of normal galaxies, which are also best
matched by the MBH densities estimated from the hard X-ray AGN LFs.
The corresponding parameters for the best match are
$(\Gamma,\epsilon)=(-0.21^{+0.28}_{-0.33},0.12^{+0.02}_{-0.01})$,
which suggests that the $\mbh-\sigma$ relation does not evolve with
redshift. The $\epsilon$ obtained by adopting the $\mbh-\sigma$
relation is larger, which is because the MBH densities at $z\sim 0$
estimated from the $\mbh-\sigma$ relation is smaller than that from
the $\mbh-\mst$ relation. Considering the $1\sigma$ errors in the
normalizations of the $\mbh-\mst$ and $\mbh-\sigma$ relations and the
AGN LF, there are the additional errors $\pm0.03$ in the estimated
$\epsilon$.  By combining these additional errors with the errors
obtained from the above $\chi^2$ fitting and averaging the $\epsilon$
obtained from the two relations, we have
$\epsilon=0.11^{+0.04}_{-0.03}$, which is consistent with the
constraint obtained in \citet{YL08}.

As seen from Figures~\ref{fig:f1} and \ref{fig:f2}, the MBH mass
density at $z\sim 0$ is about $(3.5-4)\times 10^5\msun$, which may be
slightly smaller than that obtained by others \citep[e.g.,][]{GD07,
Shankar09}. This difference is mainly due to the difference in the
normalization of the adopted $M_{\bullet}-\sigma$ relation and
different treatment on the dependence of the MBH mass density on the
Hubble constant. These differences lead to a change of the MBH
density estimates at different redshifts by the same factor, but the
shape of the MBH density evolution does not change and thus the
constraint on $\Gamma$ is not affected. 

In the above calculations of the MBH mass densities, the SMFs and VDFs
are extrapolated to the low-mass or low-velocity dispersion end, of
which part may be not accurately determined by the observations at all
the redshifts considered in this paper \citep[see][]{Ilbert10,
Bezanson11}. To see whether the final results are affected by the
extrapolation, we also set the lower limit of the MBH mass to
$10^8\msun$ and re-do the above matching. And we find
$(\Gamma,\epsilon)=(0.61^{+0.21}_{-0.20},0.18^{+0.02}_{-0.02})$ to
match the MBH mass densities estimated from the SMFs of normal
galaxies with that estimated from the AGNs, and find
$(\Gamma,\epsilon)=(-0.62^{+0.20}_{-0.20},0.19^{+0.02}_{-0.01})$ to
match the MBH mass densities estimated from the VDFs of normal
galaxies. The constraints on $(\Gamma,\epsilon)$ for MBHs with mass
$>10^8\msun$ are roughly consistent with that for MBHs with mass
$>10^6\msun$.  

\section{Discussion}\label{sec:discussion}

The positive evolution of the $\mbh-\mst$ relation found in this paper
is consistent with that found by \citet{Merloni10} (see also
\citealt{Jahnke09} and \citealt{Bennert11}), which suggests that the
growth of MBHs predates the assembly of spheroids and the spheroids
experience additional growth after the quench of their central nuclear
activities. The non-evolution of the $\mbh-\sigma$ relation found here
appears different from the positive evolution found by
\citet{Woo06,Woo08}. We note here that this  difference may be
lessened as the MBH masses estimated in \citet{Woo06} and
\citet{Woo08} may be over-estimated by a factor of two as suggested by
the uncertainties in the virial factor recently revealed by
\citet{Graham11}.  However, the exact reason for this difference is
not clear as our results are obtained through the global evolution of
the MBH mass densities, different from the way adopted in
\citet{Woo06,Woo08} for a sample of individual AGNs/QSOs.  And the
VDFs estimated by \citet{Bezanson11} are the very first estimates of
VDFs at redshift $z\neq 0$, which may suffer from various
uncertainties as discussed in \citet{Bezanson11}.  If the
non-evolution of the $\mbh-\sigma$ relation is true, nevertheless, it
suggests the velocity dispersion of galactic bulges does not increase
although the masses of the bulges can be significantly enlarged after
the quench of their central nuclear activities.

The evolution of the relations between the MBH mass and galaxy
properties are intensively investigated theoretically in the scenario
of co-evolution of galaxies and MBHs since the discovery of
these scaling relations. For those early models that adopt rapid and
strong feedback due to energy output from the central AGNs which
terminates star formation, the scaling relations are expected to
evolve little and have very small scatters
\citep[e.g.,][]{Granato04,Robertson06,diMatteo05,Springel05}.  Later
models do suggest that the $\mbh-\mst$ relation evolve positively with
redshift though with various degrees of evolution
\citep[e.g.,][]{Hopkins07,Croton06,Fontanot06, Malbon07,Lamastra10},
by considering detailed dissipation processes occurred during major
mergers of galaxies and the acquiring of bulge masses by dynamical
processes such as disk instabilities or disrupting stellar discs.
However, the expected $\mbh-\sigma$ relation is almost independent of
redshift \citep[][]{Hopkins07}. Apparently the constraints that we
obtained in this paper are roughly consistent with the theoretical
studies of \citet{Hopkins07}.  Furthermore, we note that \citet[][see
also \citealt{Oser12}]{Oser10} recently proposed a two-phase galaxy
formation scenario, in which galaxies roughly double their masses
after $z=1$ due to accretion and minor mergers while velocity
dispersion drops slightly. If MBHs mainly obtained their masses
through efficient accretion triggered by major mergers, then our
constraints are consistent with the two-phase galaxy formation
scenario.

\subsection{Mass-to-energy conversion efficiency in AGNs/QSOs}

The constraints on $\Gamma$ obtained above may depend on the use of a
constant $\epsilon$, i.e., $\epsilon$ is independent of $\mbh$,
redshift, and other physical quantities involved in the accretion
processes. We argue that the use of a constant $\epsilon$ in AGNs/QSOs
is appropriate as follows: (1) $\epsilon$ is determined mainly by the
spins of MBHs in the standard disk accretion scenario; (2) the
majority of the AGNs/QSOs are accreting via thin disks with high
Eddington ratios ($\ga 0.1$); and (3) the spins of individual MBHs can
quickly reach an equilibrium value and stay at that value for most of
the AGN lifetime as suggested by theoretical models
\citep[e.g.,][]{Volonteri05,Shapiro05, Hawley07, Maio12}, which
suggests a roughly constant $\epsilon$ at all redshift. Note that some
authors introduced a dependence of $\epsilon$ on the redshift or MBH
mass, such as \citet{DL11,MR11,Lietal12}, etc., according to the current
observations. However, these results may be only due to some
observational biases \citep[e.g.,][]{Raimundo12} and need further
investigation.  Nevertheless, we note here that one could also
introduce a cosmic evolution to $\epsilon$ before understanding the
underlying physics. For example, if we assume $\epsilon(z)
=\max\{0.057,\min[\epsilon_0(1+z)^{\kappa},0.31]\}$ but the
$\mbh-\mst$ relation does not evolve with redshift, where $0.31$ and
$0.057$ are the $\epsilon$ that an efficiently accreting MBH-disk
system could reach if the MBH spin is either $0.998$ (the maximum spin
of a MBH, see \citealt{Thorne74}) or $0$ (a Schwarzschild MBH). In
this case, an acceptable fit can also be found and the best-fit
parameters are $(\kappa,\epsilon_0)=(1.3^{+0.9}_{-0.4},
0.04^{+0.01}_{-0.02})$ for the $\mbh-\mst$ relation, which suggests
that $\epsilon$ is significantly higher at $z\sim 2$ than at $z\sim
0$. However, the MBH densities estimated from AGNs are an integration
of $\frac{d\rho_{\bullet}}{dz}$ over $z$, and
$\frac{d\rho_{\bullet}}{dz}$ is a function of $\epsilon(z)$ and has to
be determined to high redshift. The MBH densities estimated from the
SMFs in this paper only cover the redshift up to $\sim1.2$ and may
poorly constrain the evolution of $\epsilon(z)$ at higher $z$. If
$\epsilon$ is significantly higher at higher redshift, a significantly
negative evolution in the $\mbh-\sigma$ relation is required.  Future
measurements on the SMFs and VDFs at $z\ga 1.2$ may help to determine
the MBH densities at higher $z$ and thus may further help to put
constraints on whether $\epsilon$ significantly evolves with redshift.

To close the discussion on this issue, we remark here that the constraint
on the cosmic evolution of the $\mbh-\mst$ (or $\mbh-\sigma$) relation
obtained in this paper is robust if the efficiency $\epsilon$ of the
efficient accretion processes in QSOs/AGNs is roughly a constant,
which may be true as suggested by some physical models of the spin
evolution of MBHs \citep[e.g.,][]{Volonteri05,Shapiro05, Hawley07,
Maio12}.

\subsection{Intrinsic scatters in the $\mbh-\mst$ or the $\mbh-\sigma$
relation}

The intrinsic scatters in the $\mbh-\mst$  and the $\mbh-\sigma$ relations
have been assumed not to evolve with redshift in obtaining the constraints on
the cosmic evolution of the relations. If the intrinsic scatter in the
$\mbh-\mst$ relation increases significantly with increasing redshift,
the parameter $\Gamma$ can still be consistent with $0$ to match the
MBH densities estimated from normal galaxies to that from AGNs. To
settle onto the observed local $\mbh-\mst$ relation with a smaller intrinsic
scatter but with the same normalization, however, it is necessary for
those galaxies at a fixed $\mst$ with relative large MBHs to accrete
more stars and for those with relatively small MBHs not to accrete
many stars after the quenching of the nuclear activities.  This is not
likely to be the case for the stochastic increasing of $\mst$ due to
minor mergers or other dynamical processes like disk instabilities.

In addition, we note that estimation of the intrinsic scatter of the
$\mbh-\mst$ (or $\mbh-\sigma$) relation needs to determine the measurement
errors in both the MBH and stellar masses, and it is still challenging to
accurately determine these measurement errors \citep[see discussions
in][]{Graham11}. \citet{Graham12} shows that the total scatter could 
range from 0.44~dex to 0.7~dex for different types of galaxies
(see Table~1 therein).

\subsection{Alternative $\mbh-\mst$ or $\mbh-\sigma$ relation}

In the analysis in Section~\ref{sec:fit}, we adopt the single
power-law form for the $\mbh-\mst$ relation given by \citet{Haring04}.
Recently \citet{Graham12} suggests that this relation may be better
described by a broken power-law than the single power-law shown in
Equation (\ref{eq:mbhms}).  To see the effects of this new development
on the constraints obtained above, here we replace the $\mbh-\mst$
relation at $z=0$ shown in Equation (\ref{eq:mbhms}) by the broken
power-law form given in \citet{Graham12}, i.e.,
$\langle \log \mbh\rangle (\mst; z=0)= (8.38\pm
0.17) + (1.92\pm0.38)\log [\mst/(7\times 10^{10}\msun)]$ at $\mst<
7\times 10^{10}\msun$ and $(8.40\pm 0.37)+ (1.01\pm 0.52)\log
[\mst/(7\times 10^{10}\msun)]$ at $\mst\ge 7\times 10^{10}\msun$,
respectively.  Similarly, we also assume that the intrinsic scatter of
this relation is $0.3$~dex. By doing the same analysis as that in
Section \ref{sec:fit}, we obtain the constraints on
$(\Gamma,\epsilon)$ as $(0.44^{+0.74}_{-0.75}, 0.06^{+0.04}_{-0.01})$. The best fit
value of $\Gamma=0.44$ is still consistent with that obtained in
Section~\ref{sec:fit} ($0.64^{+0.27}_{-0.29}$) within $1-\sigma$ error
but its uncertainty ($^{+0.74}_{-0.75}$) is large. Compared with the constraints
on $(\Gamma,\epsilon)$ obtained for the single power-law $\mbh-\mst $
relation, the larger uncertainty of $\Gamma$ obtained here is mainly
because the relative larger uncertainties in 
the slope of the broken power-law $\mbh-\mst $ relation adopted here lead
to larger uncertainties in the estimations of the MBH mass densities
compared with that for the single power-law $\mbh-\mst$ relation
adopted above. The $\epsilon$ obtained here ($0.06$) is also
substantially smaller than that obtained for the single power-law
$\mbh-\mst$ relation, mainly because of the relatively higher
zero-point of the broken power-law given by \citet{Graham12}.

In the analysis in Section~\ref{sec:fit}, we also adopt the single
power-law form for the $\mbh-\sigma$ relation given by
\citet{Gultekin09}, which is largely consistent with those estimated
by others (see references therein).  Recently, \citet{Graham11}
updated the $\mbh-\sigma$ relation and found that this relation for
the barred galaxies may be different from that for those non-barred
galaxies/ellipticals. If assuming that the $\mbh-\sigma$ relation is
the same as that for ellipticals at $\sigma>180\kms$ [i.e.,
$\langle \log \mbh\rangle (\sigma; z=0)
= (8.22\pm 0.09)\pm (5.30\pm0.77) \log(\sigma/200\kms)$ with an
intrinsic scatter of 0.29] and the same as that for the barred
galaxies at $\sigma<180\kms$ [i.e.,
$\langle\log \mbh\rangle (\sigma;z=0)= (8.15\pm 0.06)\pm
(5.95 \pm0.44)\log(\sigma/200\kms)$ with an intrinsic scatter of $0.35$;
see Table~2 in \citealt{Graham11}], we find $(\Gamma,\epsilon)= 
(-0.86^{+0.31}_{-0.30},0.14^{+0.02}_{-0.01})$. The constraint obtained
here for $\Gamma$ is substantially different from that obtained in
Section~\ref{sec:fit}, which suggests a significant negative evolution
of the velocity dispersion of individual big galaxies, i.e., the velocity
dispersions of big galaxies decrease by $\sim 20\%$, which is marginally
compatible with the hierarchical galaxy formation scenario as recently proposed
by \citet[e.g.,][]{Oser10,Oser12}.
However, the $\epsilon$ obtained here is slightly higher than that
obtained above for a single power-law $\mbh-\sigma$ relation, which is
mainly due to the smaller intrinsic scatter of the adopted $\mbh
-\sigma$ relation and the smaller normalization for the relation at
$\sigma <180\kms$. Both of those factors lead to slightly smaller MBH mass
densities in all the redshift bins and hence slightly higher $\epsilon$.

\section{Conclusions}
\label{sec:conclusion}

In this paper, we study the cosmic evolution of the $\mbh-\mst$
relation and the $\mbh-\sigma$ relation by a global method,
independent of individual MBH mass estimates. We have estimated the
cosmic evolution of MBH mass densities over the redshift range of
$z\sim 0-1.2$. The MBH mass densities are estimated from both the
SMFs/VDFs of normal galaxies determined by recent observations using
the $\mbh-\mst$ and $\mbh-\sigma$ relations, and the AGN X-ray LFs
according to the simple \citet{Soltan82} argument. By matching the MBH
densities estimated from the normal galaxies with that from the AGN
X-ray LFs, we obtain global constraints on the evolution of the
$\mbh-\mst$ relation and the $\mbh-\sigma$ relation. We find that the
$\mbh-\mst$ relation evolves with redshift positively, i.e., $\propto
(1+z)^{\Gamma}$ and $\Gamma =0.64^{+0.27}_{-0.29}$, though the
significance level is not high; however, a $\Gamma$ larger than
$1.6$ is excluded at the $3\sigma$ level. We also find that
the $\mbh-\sigma$ relation
appears not to positively evolve with redshift
($\Gamma=-0.21^{+0.28}_{-0.33}$). 
Our results suggest that the MBH
mass growth precedes the bulge mass growth but the galaxy velocity
dispersion does not increase with the mass growth of the bulge after
the quench of nuclear activity, which is roughly consistent with the
two-phase galaxy formation scenario proposed by \citet{Oser12} in
which a galaxy roughly double its masses after $z=1$ due to accretion
and minor mergers while its velocity dispersion drops slightly.

\acknowledgements
We thank the referee for helpful comments. This work was supported in
part by the National Natural Science Foundation of China under nos.\
10973001, 10973017, 11033001, and by the Bairen program from the
National Astronomical Observatories, Chinese Academy of Sciences.

\end{document}